\documentclass[conference]{IEEEtran}
\usepackage{amssymb}
\usepackage{mathrsfs}
\usepackage{amsmath}
\usepackage{latexsym}
\usepackage{graphicx}
\usepackage{bbding}
\usepackage{indentfirst}
\IEEEoverridecommandlockouts

\begin{document}
\title{Distributed Compressed Wideband Sensing in Cognitive Radio Sensor Networks}
\author{\authorblockN{Huazi~Zhang$^{1}$, Zhaoyang~Zhang$^{1}$, Yuen Chau$^{2}$
\\1. Dept.of Information Science and Electronic Engineering, Zhejiang University, China.
\\2. Singapore University of Technology and Design (SUTD), Singapore.
\\Email: ning\_ming@zju.edu.cn, yuenchau@sutd.edu.sg
}} \maketitle

\begin{abstract}
A novel distributed compressed wideband sensing scheme for Cognitive
Radio Sensor Networks (CRSN) is proposed in this paper. Taking
advantage of the distributive nature of CRSN, the proposed scheme
deploys only one single narrowband sampler with ultra-low sampling
rate at each nodes to accomplish the wideband spectrum sensing.
First, the practical structure of the compressed sampler at each
node is described in detail. Second, we show how the Fusion Center
(FC) exploits the sampled signals with their spectrum
randomly-aliased to detect the global wideband spectrum activity.
Finally, the proposed scheme is validated through extensive
simulations, which shows that it is particularly suitable for CRSN.

\end{abstract}

\section{Introduction}

Many advanced sensor network applications require that the sensed
information are delivered in the form of broadband multimedia by
resource-constrained sensor nodes, or need the sensor nodes to be
survivable in highly dynamic spectrum environments. This is hard to
achieve in traditional wireless sensor networks (WSN), since they
usually compete for limited available bandwidth on fixed unlicensed
bands. A promising sensor networking solution for these applications
is the newly introduced paradigm of Cognitive Radio Sensor Networks
(CRSN) \cite{CRSN,Zhang-2011ICC-JSCS,CWSN,CWSNsurvey}, which incorporates cognitive radio (CR)
capability into the traditional WSN. Without causing harmful
interference to the primary user (PU) systems, CRSN can provide
dynamic spectrum access and the bandwidth as high as possible to
meet the application-specific QoS requirement.


According to the NSF Spectrum Occupancy Measurements
\cite{NSFReport}, the spectrum occupation pattern is sparse.
Exploiting this sparse prior, several compressed wideband sensing
schemes are developed based on the literatures of compressed sensing
(CS) \cite{CS2}. Tian developed a distributed compressed spectrum
sensing approach for wideband CR networks \cite{Tian-CS}. Her scheme
can achieve high-performance at low sampling rate below the Nyquist
rate. Several other works \cite{CR-CS1}-\cite{CR-CS3} have also
studied the application of CS on wideband spectrum sensing. However,
to the best of our knowledge, their emphasis is not on the specific
sampling methods, i.e. little information is provided about the
structure of the sampler and their feasibility in practical
implementation. Moreover, their sampling rate for individual CR
should exceed the sum of the active bandwidth, which is still too
high for the energy-constrained CRSN nodes.

In this paper, we propose a novel distributed compressed wideband
sensing scheme for CRSN. The main contributions of this paper are
two-fold. First, we propose practical wideband sampling structure in
fine detail. Second, the required sampling rate in our scheme is
equivalent to the bandwidth of a single subband, much lower than
existing schemes. Our sampling structure is similar to the
Sub-Nyquist sampler proposed recently by Mishali and Eldar - the
modulated wideband converter (MWC) \cite{MWC1}. The major difference
is that we only deploy one sampling channel in a single CRSN node,
and each node randomly mixes the signals of all subbands by spectrum
aliasing. This simple structure is extremely convenient for
implementation in the resource-constrained CRSN, due to its low
hardware cost and power consumption.


The rest of this paper is organized as follows. In Section II, we
establish basic models for the PU signal and the CRSN, and formulate
our target problem. In Section III, we introduced the distributed
compressed wideband sensing scheme in detail. Then, extensive
simulation results are presented in Section IV to further validate
our proposed algorithm. Finally, the whole paper is concluded in
Section V.

\section{System Model and Problem Formulation}

We consider the radio environment to be an ultra-wide frequency band
where different PU systems and large numbers of CRSN nodes coexist.

\subsection{Primary User Signal}
The entire $W$ wide spectrum is centered at zero and can be divided
into $L$ non-overlapping subbands of equal bandwidth $B$, thus $W =
L \cdot B$. We assume there are $J$ PUs occupying some of the
subbands. According to \cite{NSFReport}, the PU signal is sparse
within the range of the whole spectrum. Therefore, we can safely
assume that $J$ is much smaller than $L$. Note that we will take
this assumption as our basic prerequisite throughout this paper.

We assume real-valued continuous-time wideband signal of the primary
users to be $x\left( t \right)$, which is band-limited within
$\left( { - \frac{W}{2},\frac{W}{2}} \right)$ and its Fourier
transform can be defined by
\begin{equation}
X\left( f \right) = \int_{ - \infty }^{ + \infty } {x\left( t
\right)} e^{ - j2\pi ft} {\rm{d}}t
\end{equation}

As mentioned above, the frequency domain is zero centered and
divided into $L$ subbands of equal bandwidth $B$, $X\left( f
\right)$ can be re-written as

\begin{equation}
X\left( f \right) = \sum\limits_{l =  - L_0 }^{ + L_0 } {X_l \left(
f \right)}
\end{equation}

where $L = 2L_0  + 1$ and

$X_l \left( f \right) = \left\{ {\begin{array}{*{20}c}
   {X\left( f \right),f \in \left( {\left( {l - \frac{1}{2}} \right)B,\left( {l + \frac{1}{2}} \right)B} \right)}  \\
   {0,\quad\quad\quad\quad\quad\quad\quad {\rm{otherwise}}}  \\
\end{array}} \right.$ is the signal in the $l$th subband of the wideband
spectrum. Thus, we can represent the primary signal using a vector $
{\bf{X}}\left( f \right) = \left[ {X_{_{  L_0 } } \left( f \right),
\cdots ,X_0 \left( f \right), \cdots ,X_{_{ - L_0 } } \left( f
\right)} \right]^T$. Since $x\left( t \right)$ is real-valued
continuous-time signal, the real part of Fourier transform $X\left(
f \right)$ is an even function, and the imaginary part is an odd
function. Then, obviously $ \left| {X_l \left( f \right)} \right| =
\left| {X_{ - l} \left( f \right)} \right|$.


\subsection{Secondary User System: CRSN nodes and Fusion Center}

The Secondary User (SU) system is comprised of CRSN nodes and Fusion
Centers. The spatially distributed CRSN nodes can be wireless
terminals such as webcams, cellphones and laptops. Their
application-specific QoS requires that these CRSN nodes deliver
event features reliably and timely in the form of multimedia,
resulting in high bandwidth demands. Therefore, our objective is to
design a resource-efficient wideband spectrum sensing scheme for the
CRSN.

\begin{figure}[h] \centering
\includegraphics [width=0.40\textwidth] {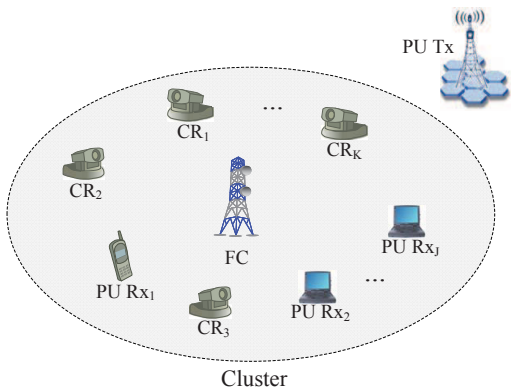}
\caption {An overview of the considered system model.} \label{Fig1}
\end{figure}

As shown in Fig.\ref{Fig1}, CRSN nodes located in a restricted
locality form a cluster \cite{Cluster-CRSN}. All nodes in one cluster sense the wide
band cooperatively. There is a Fusion Center (FC) in each cluster,
which is a special kind of node equipped with more or renewable
power sources. As a result, FC have additional computation
capability and longer transmission ranges. Since all the CRSN nodes
in a cluster are close to each other, we assume that a common
control channel can be found to exchange spectrum sensing
information.

Suppose that there are $K$ CRSN nodes engaged in the wideband
spectrum sensing task. Because these $K$ nodes are close to each
other, the PU activities in their ambient radio environment are
assumed to be identical. However, due to different channel fadings,
the received primary signal can be different at each CRSN nodes.
During the sensing interval, the primary signal received by the
$k$th CRSN nodes can be expressed by
\begin{equation}
X^k \left( f \right) = X\left( f \right)H^k \left( f \right) + W^k
\left( f \right)
\end{equation}


 Using ${\bf{H}}^k\left( f \right) = \left[ {H_{_{  L_0 } }^k \left( f \right), \cdots ,H_{_0 }^k \left( f \right), \cdots ,H_{_{-L_0 } }^k \left( f \right)} \right]^T
$ and ${\bf{W}}^k\left( f \right) = \left[ {W_{_{  L_0 } }^k \left(
f \right), \cdots ,W_{_0 }^k \left( f \right), \cdots ,W_{_{-L_0 }
}^k \left( f \right)} \right]^T $ to represent channel fading and
noise of all subbands, we can rewrite (3) in the matrix form

\begin{equation}
{\bf{X}}^k \left( f \right) = diag\left( {{\bf{H}}^k \left( f
\right)} \right){\bf{X}}\left( f \right) + {\bf{W}}^k \left( f
\right)
\end{equation}

\section{Distributed Compressed Wideband Spectrum Sensing Scheme}

In this section, we propose a cooperative spectrum sensing scheme
that can effectively distribute the sensing task to many CRSN nodes
and relieve the sensing and processing burden on a single node. We
first describe the compressed aliasing sampling scheme originated
from the Modulated Wideband Converter (MWC), then explain the fusion
strategy and corresponding recovering algorithm.

\begin{figure}[h] \centering
\includegraphics [width=0.50\textwidth] {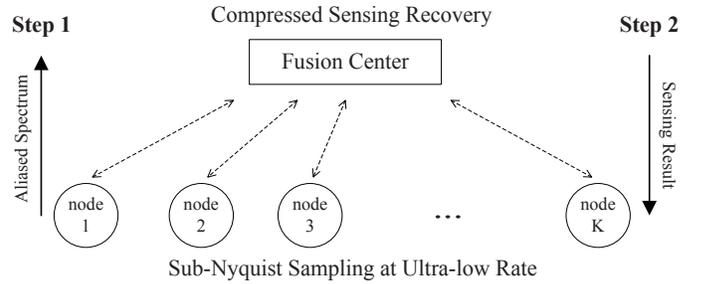}
\caption {Distributed Compressed Wideband Sensing Scheme.}
\label{Fig2}
\end{figure}

As shown in Fig.2, our proposed sensing scheme comprises two steps.

\textbf{Step 1:} $K$ CRSN nodes sample the wideband independently at
ultra-low rate equivalent to the bandwidth of a single subband, then
report the aliased spectrum information to the FC.

\textbf{Step 2:} After receiving all the aliased spectrum
information from the $K$ CRSN nodes, the the FC performs compressed
sensing recovery algorithm, and broadcast the sensing result to all
the CRSN nodes within the cluster.

\subsection{Distributed Sampling based on Spectrum Aliasing}

The sampling method is similar to the Modulated Wideband
Converter(MWC) \cite{MWC1}\cite{MWC2}. The difference is that we
only deploy one channel of the MWC system in a single CRSN node. The
sampling structure in the $k$th CRSN node is shown in Fig.3.

\begin{figure}[h] \centering
\includegraphics [width=0.50\textwidth] {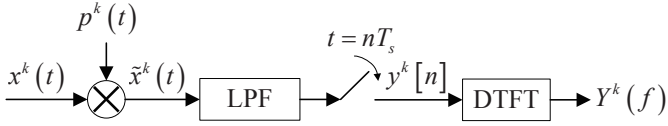}
\caption {Compressed sampling structure based on aliasing for a CRSN
node.} \label{Fig3}
\end{figure}

According to Fig.3, after the primary signal $x^k (t)$ arrives at
the $k$th CRSN node, it is multiplied by a high frequency mixing
function $p^k (t)$. $p^k (t)$ is a $T_s$-length periodic
spread-spectrum mixing function which aims to alias the spectrum and
thus obtain a mixture of signals from all the subbands.
\begin{equation}
p^k (t) = \alpha _{km} ,{\rm{     }}m\frac{{T_s }}{M} \le t \le (m +
1)\frac{{T_s }}{M},{\rm{      }}0 \le m \le M - 1
\end{equation}
where $\alpha _{km}  \in \left\{ { + 1, - 1} \right\}$, and $ p^k
(t) = p^k (t + nT_s )$ for every $n \in \mathbb{Z}$. In practice,
$p^k (t)$ is characterized by pseudorandom sequences generated by
certain seeds known at the FC. Therefore, the aliasing pattern of
the spectrum is known at the FC.

Then, the mixed analog spectrum signal goes through a Low-Pass
Filter (LPF) with cutoff frequency $1/2T_s$ and the filtered signal
is sampled at rate $f_s=1/T_s$. For convenient spectrum access, we
choose $f_s$ to be equal to $B$, so low that existing A/D converters
are competent for the task.

We perform discrete-time Fourier transform (DTFT) on the sampled
data, in order to obtain the frequency domain form of the mixed
spectrum. According to \cite{MWC1},
\begin{equation}
Y^k (e^{j2\pi fT_s } ) = \sum\limits_{l =  - L_0 }^{ + L_0 } {c^k_l
} X^k (f - lB )
\end{equation}

where $ f \in \mathcal{F}_s=\left[ { - B /2, + B /2} \right]$, $L_0
= \left\lceil {\frac{{W - B }}{{2B }}} \right\rceil $, and $ c^k_l =
B\int_0^{1/B } {p^k (t)e^{ - j\frac{{2\pi }}{{T_s }}lt} } dt $ is
the Fourier coefficient of $ p^k (t)$. After a closer examination,
we can easily find that (6) is a weighted sum of the spectrum of all
subbands, and can be rewritten in the matrix form:
\begin{equation}
Y^k \left( f \right) = {\bf{c}}^k \left( {diag\left( {{\bf{H}}^k
\left( f \right)} \right){\bf{X}}\left( f \right) + {\bf{W}}^k
\left( f \right)} \right)
\end{equation}
where ${\bf{c}}^k  = \left[ {c^k_{ L_0 } , \cdots ,c^k_{ - L_0 } }
\right]$.



At last, we calculate the average level of the band-limited spectrum
mixture, and transmit this aliased spectrum information to the FC
through the control channel.
\begin{equation}
Y^k  \equiv \bar Y^k (f) = \int\limits_{ - B/2}^{ + B/2} {\left|
{Y^k (f)} \right|df}
\end{equation}

\subsection{Fusion Strategy and Recovering Algorithm}

\begin{figure*}[!t]
\normalsize
\setcounter{equation}{10}
\begin{equation}
\begin{array}{l}
 {\bf{A}}_{K \times L}  = \left( {\begin{array}{*{20}c}
   {\alpha _{1,0} } &  \ldots  & {\alpha _{1,L - 1} }  \\
    \vdots  &  \ddots  &  \vdots   \\
   {\alpha _{K,0} } &  \cdots  & {\alpha _{K,L - 1} }  \\
\end{array}} \right)\left( {\begin{array}{*{20}c}
   {\theta ^{L_0  \cdot 0} } &  \ldots  & {\theta ^{ - L_0  \cdot 0} }  \\
    \vdots  &  \ddots  &  \vdots   \\
   {\theta ^{L_0  \cdot (L - 1)} } &  \cdots  & {\theta ^{ - L_0  \cdot (L - 1)} }  \\
\end{array}} \right)\left( {\begin{array}{*{20}c}
   {d_{L_0 } } &  \ldots  & 0  \\
    \vdots  &  \ddots  &  \vdots   \\
   0 &  \cdots  & {d_{ - L_0 } }  \\
\end{array}} \right)\left( {\begin{array}{*{20}c}
   {H_{L_0 } } &  \ldots  & 0  \\
    \vdots  &  \ddots  &  \vdots   \\
   0 &  \cdots  & {H_{ - L_0 } }  \\
\end{array}} \right) \\
 {\rm{   }} = {\bf{S}}_{K \times L}  \times {\bf{F}}_{L \times L}  \times {\bf{D}}_{L \times L}  \times {\bf{H}}_{L \times L}  \\
 \end{array}
\end{equation}

\hrulefill
\end{figure*}
\setcounter{equation}{8}

$K$ CRSN nodes engage in the spectrum sensing task and report their
spectrum mixture value to the fusion center, all the reported value
can compose a $K$-length vector ${\bf{Y}} = \left[ {Y^1 , \cdots
,Y^K } \right]^T$

Here, we further use the assumption that all the $K$ CRSN nodes and
the FC are close to each other and their channel fadings are
approximately the same, i.e. ${\bf{H}}^1 \left( f \right) = \cdots =
{\bf{H}}^k \left( f \right) = {\bf{H}}\left( f \right)$. It can be
estimated at the FC and broadcasted to all the CRSN nodes.

The $K$-length vector collected from $K$ distributed CRSN nodes to
the FC can be expressed as
\begin{equation}
\begin{array}{l}
 {\bf{Y}} = \int\limits_{ - B/2}^{ + B/2} {\left| {{\bf{C}}\left( {diag\left( {{\bf{H}}\left( f \right)} \right){\bf{X}}\left( f \right) + {\bf{W}}\left( f \right)} \right)} \right|df}  \\
  = \int\limits_{ - B/2}^{ + B/2} {\left| {{\bf{A}}\left( f \right){\bf{X}}\left( f \right) + {\bf{W'}}\left( f \right)} \right|df}  \\
 \end{array}
\end{equation}
where $\bf{C}$ is the Fourier coefficient matrix of mixing functions
$ {\bf{p}} \left( t \right) $, ${\bf{A}}\left( f \right)$ is a $K
\times L$ measurement matrix, ${\bf{A}}_{kl} \left( f \right) =
c^k_l H_l \left( f \right) $ and ${\bf{W'}} \left( f \right)$ is
also a noise vector.

Assuming that the channel fading are also almost flat within a
single subband, i.e. $H_l  \equiv \bar H_l (f) = \int\limits_{ -
B/2}^{ + B/2} {\left| {H_l (f)} \right|df}$, equation (9) can be
rewritten as

\begin{equation}
{\bf{Y}} = {\bf{AX}} + {\bf{W'}}
\end{equation}
where $ {\bf{A}}_{kl}  = c^k_l H_l$, and we denote ${\bf{X}} =
\int\limits_{ - B/2}^{ + B/2} {\left| {{\bf{X}}\left( f \right)}
\right|df} $, ${\bf{W'}} = \int\limits_{ - B/2}^{ + B/2} {\left|
{{\bf{W'}}\left( f \right)} \right|df} $ to be the average level of
the primary signal and noise, respectively.

Based on the analysis in \cite{MWC1}, $\bf{A}$ can be expressed as
(11) (on the top of next page), where $\theta = e^{ - j\frac{{2\pi
}}{L}}$ and \setcounter{equation}{11}
\begin{equation}
d_l = \frac{1}{{T_s }}\int_0^{\frac{{T_s }}{L}} {e^{ - j\frac{{2\pi
}}{{T_s }}lt} } dt = \left\{ {\begin{array}{*{20}c}
   {\frac{1}{L}{\rm{        }}l = 0}  \\
   {\frac{{1 - \theta ^l }}{{j2\pi l}}{\rm{      }}l \ne 0}  \\
\end{array}} \right.
\end{equation}

Considering (10), where $\bf{X}$ is is an unknown sparse vector of
dimension $L$, $\bf{Y}$ is the measured vector of dimension $K$, and
$\bf{A}$ is the known random measurement matrix of size $K \times L$
. Our final goal is to recover the original primary signal $\bf{X}$
from the measured compressed vector $\bf{Y}$ and determine the idle
subbands for opportunistic spectrum access. This is exactly the
compressed sensing (CS) problem \cite{CS2} since $K<<L$.

%

It is proved that a random sign matrix, whose entries are drawn
independently from $\pm1$ with equal probability, has the RIP of
order $S$ if $K \ge C \cdot S\log \left( {L/S} \right)$
\cite{MWC2}\cite{CS2}, where $C$ is a positive constant. Also, any
fixed unitary row transformation of a random sign matrix has the
same RIP \cite{MWC1}. From (12), we can see ${\bf{S}}_{K \times L}$
is a random sign matrix, this implies that ${\bf{A}}_{K \times L}$
also has RIP and (10) can be solved using existing CS recovery
methods.

In this paper, we employ the basis pursuit (BP), a convex
programming method that can solve CS problems in polynomial-time. We
choose BP for the following two reasons. First, the sparsity level
of the input signal is known in other OMP \cite{CS_OMP} related CS
recovery algorithm. However, in our scenario, the sparsity level
means the number of the active subbands and can not be estimated
beforehand. Second, the BP algorithm has robust performance in the
noisy scenario.

We first estimate global spectrum vector ${{\bf{\hat X}}}$ through
${l_1}$ minimization based noisy BP algorithm:
\begin{equation}
\min \left\| {{\bf{\hat X}}} \right\|_{l_1 }\ \ \ {\rm{subject}\
{to}}\left\| {{\bf{A\hat X - Y}}} \right\|_{l_2 }  \le \sigma _W^2
\end{equation}
where $\sigma _W^2$ is the variance of the Gaussian noise ${W'}$.

And then define a decision vector ${{\bf{\hat d}}}$ of the PU
activity state by comparing the $l$th subband estimation ${\hat
X}_l$ with a decision threshold $\lambda$:

\begin{equation}
{\bf{\hat d}} = \left[ {\hat d_{L_0 } , \cdots ,\hat d_l , \cdots
\hat d_{ - L_0 } } \right]
\end{equation}
where $ \hat d_l  = \left\{ {\begin{array}{*{20}c}
   {0,\hat X_l  < \lambda }  \\
   {1,\hat X_l  > \lambda }  \\
\end{array}} \right.$


The decision vector serves as our compressed wideband spectrum
sensing result. After this decision vector is obtained at FC, this
final result is broadcasted to all the CRSN nodes within the
cluster. The CRSN nodes then access the spectrum according to this
sensing result.

\section{Performance Evaluation}

To evaluate the accuracy of our proposed compressed sensing recovery
algorithm, we define the normalized root mean-square estimation
error (MSE):

\begin{equation}
{\rm{MSE}} = {\mathop{\rm E}\nolimits} \left( {\frac{{\left\|
{{\bf{\hat X}} - {\bf{X}}} \right\|_{l_2 } }}{{\left\| {\bf{X}}
\right\|_{l_2 } }}} \right)
\end{equation}

Just like any cognitive radio network, we choose the probability of
detection $P_d$ and the probability of false alarms $P_{f}$ as our
wideband sensing performance metrics. In the multi-band scenario,
they can be expressed as

\begin{equation}
\begin{array}{l}
 P_d  = {\mathop{\rm E}\nolimits} \left( {\frac{{\mathop {Num}\limits_{l = L_0 ,..., - L_0 } \left( {d_l  = 1\ {\rm{and}}\ \hat d_l  = 1} \right)}}{{\mathop {Num}\limits_{l = L_0 ,..., - L_0 } \left( {d_l  = 1} \right)}}} \right) \\
 P_{f}  = {\mathop{\rm E}\nolimits} \left( {\frac{{\mathop {Num}\limits_{l = L_0 ,..., - L_0 } \left( {d_l  = 0\ {\rm{and}}\ \hat d_l  = 1} \right)}}{{\mathop {Num}\limits_{l = L_0 ,..., - L_0 } \left( {d_l  = 0} \right)}}} \right) \\
 \end{array}
\end{equation}
where $d_l  = \left\{ {\begin{array}{*{20}c}
   {0,\ {\rm{the }}\ l{\rm{th\  subband\  is\  idle}}}  \\
   {1,\ {\rm{the }}\ l{\rm{th\  subband\  is\  busy}}}  \\
\end{array}} \right.$ represents the PU activity pattern on the
spectrum.

In the following experiments, we set the system parameters to be an
ultra-wideband regime. For all scenarios, we set the total bandwidth
of the spectrum $W$ to be $6$GHz, and it is equally divided into
$201$ subbands of bandwidth $30$MHz. Among them, $30$ subbands are
occupied by PUs. Note that the spectrum is symmetric to the zero
point, there are $J=15$ independent active PUs, half the number of
the occupied subbands. The PU occupation ratio is $15$\%.

\emph{1) Comparison of Sampling Rate:}

In Table I, we compare the sampling rate between our scheme and
existing ones. Here, we explain how our scheme achieves same or
better performance. In our setting, for Nyquist Sampling, the
sampling rate should be twice of the one-sided bandwidth, which is
$6$GHz. For existing compressed spectrum sensing schemes
\cite{Tian-CS}-\cite{CR-CS3}, each node first obtains local sensing
result. Thus, the dimension of the local measured vector should be
about $4\times$ the sparsity level to achieve exact recovery. This
is the four-to-one practical rule well known to many CS researchers
\cite{CS2}. As a result, the sampling rate should be four times of
the occupied bandwidth, which is $3.6$GHz in our setting.

We should point out that although the required sampling rate
decreased a lot with previous compressed spectrum sensing methods,
it is still too high for the resource-constraint CRSN in the
ultra-wideband regime. The novel idea of our scheme is to find a
distributive compressed sampling structure, and further reduce the
individual sampling rate. In our scheme, we effectively distribute
the sensing load evenly to each CRSN nodes by splitting the
compressed sampling process, and acquire the sensing result at the
FC. Our individual sampling rate is $f_s$ and the sum sampling rate
for a cluster is $K \times f_s$. With the same CS recovery
algorithm, the performance is mainly influenced by the sum sampling
rate. In other words, the performance remains unchanged as long as
$K \times f_s$ is a constant. The decrease in $f_s$ will result in
the increase of $K$, and vice versa. We can allocate individual
sampling rates that the sum sampling rate of all CRSN nodes reaches
four times of the occupied bandwidth. When $K \times f_s=3.6$GHz,
our scheme can achieve equivalent performance to existing compressed
spectrum sensing schemes. For convenience of spectrum access, we
choose the local sampling rate to be equal to the bandwidth of SU
subband, i.e. $f_s=B=30$MHz. As more CRSN nodes engage in the
sensing task, more robust performance can be achieved.

For all the following Monte Carlo experiments, we repeat a hundred
thousand times to compute the target value.


\begin{table}[!t]
\renewcommand{\arraystretch}{1.1}
\caption{A Comparison of Sampling Rate Required by Different
Schemes} \label{table_example} \centering
\begin{tabular}{|c|c|}
\hline
\multicolumn{2}{|c|}{System Parameters} \\
\hline
Total Bandwidth & 6 GHz \\
\hline
Subband Bandwidth & 30 MHz\\
\hline
Occupied Bandwidth & 900 MHz \\
\hline
Total Subbands & 201 \\
\hline
Occupied Subbands & 30 \\
\hline
Occupation Ratio & 15\% \\
\hline\hline
\multicolumn{2}{|c|}{Comparison of Sampling Rate at Individual Node} \\
\hline
Conventional Nyquist Sampling & 6 GHz \\
\hline
Existing Compressed Spectrum Sensing & 3.6 GHz\\
\hline
Distributed Compressed Wideband Sensing & 30 MHz \\
\hline
\end{tabular}
\end{table}

\emph{2) Performance of Compressed Sensing}

In the first experiment, we examine the effect of measurement noise
and $K$ on the recovery accuracy in terms of MSE. As shown in Fig.4,
we adjust the $K$ from $25$ to $50$, and find that the MSE drops as
$K$ increases. It shows that, if too few measurements are collected,
the performance of the compressed sensing algorithm will
deteriorate.

\begin{figure}[h] \centering
\includegraphics [width=0.43\textwidth] {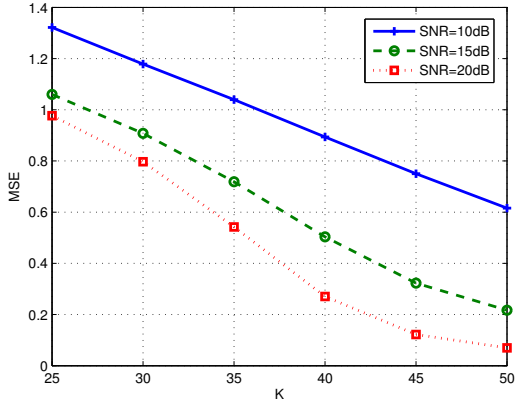}
\caption {MSE of compressed sensing vs $K$.} \label{Fig4}
\end{figure}


\emph{3) Performance of Distributed Wideband Sensing}

In the second experiment, we evaluate the performance of distributed
wideband sensing in terms of $P_d$ and $P_f$ versus the number of
engaged CRSN nodes $K$. We adjust $K$ from $25$ to $50$,  and find
that $P_d$ increases quickly, while $P_f$ drops. This implies that
as more CRSN nodes engaged in the sensing task, the detecting
performance of the cluster improves. The numerical results are shown
in Fig.5.

\begin{figure}[h] \centering
\includegraphics [width=0.43\textwidth] {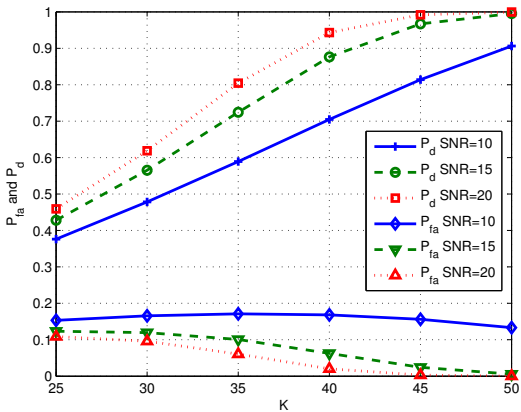}
\caption {Probability of detection and false alarm vs $K$.}
\label{Fig6}
\end{figure}

We also computed the Receiver Operating Characteristic (ROC) curve
under different $K$. As shown in Fig.6, the closer the curve is to
the upper left corner, the better the performance is. The four
curves represent the cases when CRSN nodes numbers are $25$, $30$,
$40$ and $60$. The performance improves as the engaging CRSN nodes
number $K$ grows. Interestingly, we observe that as $K$ outnumbers
$J$ by four times, the detector's performance is nearly optimal.

\begin{figure}[h] \centering
\includegraphics [width=0.43\textwidth] {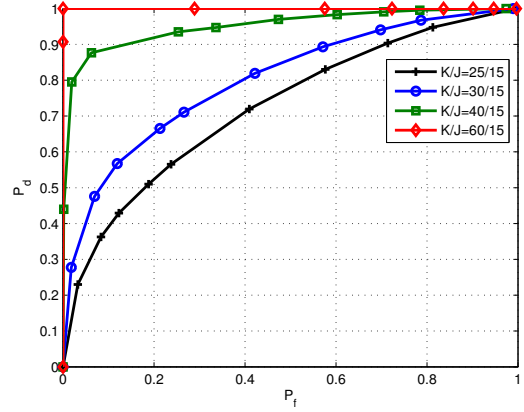}
\caption {ROC curve of the proposed detector.} \label{Fig7}
\end{figure}

\section{Conclusion}

In this paper, we proposed a distributed compressed wideband sensing
scheme for the resource-constrained Cognitive Radio Sensor Networks
(CRSN). First, we described the considered scenario by modeling the
primary user signal and the clustered CRSN structure. Second, we
provided a specific and practical structure for the sampler, which
has very low sampling rate at individual node. Then, we introduced
the sensing information fusion scheme, and described the compressed
sensing recovery algorithm at the fusion center. Finally, we carried
out several numerical simulations to validate our proposed scheme.

\section{Acknowledgement}
This work was supported by the National Hi-Tech Research
and Development Program (863 program) of China (No. 2007AA01Z257),
National Basic Research Program (973 Program) of China (No.
2009CB320405) and National Natural Science Foundation Program of
China (No. 60802012, 60972057, 61001098). It is also partly
supported by the Singapore University of Technology and Design (No.
SRG-EPD-2010-005).

\end{document}